\documentclass[preprint,prl,superscriptaddress]{revtex4}

\usepackage{amsmath} % American Mathematical Society Package

\usepackage{latexsym,amssymb} % additional LaTeX symbols

\usepackage{graphicx} % graphics in LaTeX
\usepackage{epsfig} % eps figures
\usepackage{siunitx}

\begin{document}

% newcommands
\newcommand{\ud}{\mathrm{d}} % integral d
\newcommand{\lL}{\ell}
\newcommand{\lone}{(\ell+1)}
\newcommand{\llone}{\ell(\ell+1)}
\renewcommand\arraystretch{1.1}% set row height in tabular ( value=1.0 is for standard spacing )
\newcommand\grwdth{0.95\columnwidth} % common width of all graphs 0.75 is ok, but it can be smaller
\newcommand{\BO}[1]{\mathcal{O}\left( #1\right)} %% Landau Symbols - big/small O notation

\title{ Equivariant wave maps exterior to a ball}

\author{Piotr Bizo\'n}
\affiliation{Institute of Physics, Jagiellonian University, Krak\'ow, Poland}
\author{Tadeusz Chmaj}
\affiliation{Institute of Nuclear
   Physics, Polish Academy of Sciences,  Krak\'ow, Poland}
   \affiliation{Cracow University of Technology, Krak\'ow,
    Poland}
\author{Maciej Maliborski}
\affiliation{Institute of Physics, Jagiellonian University, Krak\'ow, Poland}

\date{\today}
\begin{abstract}
We consider the exterior Cauchy-Dirichlet problem for equivariant wave maps from $3+1$ dimensional Minkowski spacetime into the three-sphere. Using mixed analytical and numerical methods we show that, for a given topological degree of the map, all solutions starting from smooth finite energy initial data converge to the unique static solution (harmonic map). The asymptotics of this relaxation process is described in detail.
We hope that  our  model will provide an attractive mathematical setting  for gaining insight into dissipation-by-dispersion phenomena, in particular the soliton resolution conjecture.
\end{abstract}

\maketitle
\section{Introduction}
Dissipation of  energy by dispersion is the key mechanism of relaxation to a static equilibrium in infinite dimensional Hamiltonian systems on unbounded domains \cite{soffer}.
 It occurs in a variety of physical situations ranging from dynamics of gas bubbles in a compressible fluid \cite{sw} to formation of black holes in gravitational collapse \cite{ks}.
Despite great physical importance, mathematical understanding of dissipation by dispersion phenomena is still in its infancy, especially in the non-perturbative regime. Thus, to make progress it would be useful to have a good toy-model of this phenomenon. Among the desired features of such a model we list the following: 1) the evolution equation is as simple as possible, preferably a scalar semilinear equation, 2)
 solutions are globally regular in time for all reasonable initial data, so that one need not to worry about the development of singularities, 3) the static equilibrium is unique (so that there is no competition between different attractors) and rigid (so that no modulation analysis is needed), 4) the linearization about the static solution has a purely continuous spectrum (no unstable, neutral or oscillatory modes), 5) the linearized waves  decay fast enough to start the Duhamel iteration argument (the main technical tool in proving asymptotic stability).
 If  all these properties hold, then it is natural to expect that the static solution is a global attractor, \emph{i.e.},  all solutions converge to it as time goes to infinity while the excess energy is being radiated away to infinity. This is a particularly simple case of the soliton resolution conjecture which asserts
 that generic global in time solutions of nonlinear dispersive equations resolve asymptotically into a superposition of a coherent structure (solitons)  and free radiation \cite{soffer,tao}.

 The purpose of this paper is to propose a model which enjoys all the properties listed above. We hope that the insight gained by studying this model may help in understanding real physical problems.
 The rest of the paper is organized as follows. In section~2 we define the model. Static solutions are discussed in section~3. In section~4 we describe the spectral properties of the linearized problem. Finally, these analytic results are confronted with numerical simulations in section~5.
\section{Model}
 A map $U:\mathcal{M}\mapsto \mathcal{N}$  from a spacetime $(\mathcal{M},g_{\alpha\beta})$ into a Riemannian manifold $(\mathcal{N},G_{AB})$ is said to be the wave map if it is a critical point of the action
\begin{equation}
  \label{action}
  S[U]:= \int_{\mathcal{M}}
  \frac{\partial X^A}{\partial x^{\alpha}}
  \frac{\partial X^B}{\partial x^{\beta}}
  \;g^{\alpha\beta}\:G_{AB}\sqrt{-g}\,dx\,,
\end{equation}
where $x^{\alpha}$ and $X^A$ are local coordinates on $\mathcal{M}$ and
$\mathcal{N}$, respectively. Variation of the action \eqref{action} gives a system of semilinear wave equations
\begin{equation}
  \label{WM}
  \square_{g}X^C + \Gamma^{C}_{AB}\frac{\partial X^A}{\partial x^{\alpha}}
  \frac{\partial X^B}{\partial x^{\beta}}g^{\alpha\beta} = 0,
\end{equation}
where $\square_{g}:= \frac{1}{\sqrt{-g}} \partial_{\alpha} \left(g^{\alpha\beta} \sqrt{-g} \, \partial_{\beta}\right)$ is
the wave operator associated with the metric $g_{\alpha\beta}$ and $\Gamma^{C}_{AB}$ are the
Christoffel symbols of the target metric $G_{AB}$.  In this paper we consider the
wave maps from
$\mathcal{M}=\mathbb{R}^{3+1}$,  the $3+1$ dimensional Minkowski spacetime, into $\mathcal{N}=S^3$, the unit 3-sphere.  On $\mathbb{R}^{3+1}$ we use standard
spherical coordinates $(t,r,\theta,\phi)$ and on $S^3$
we choose polar coordinates $X^A=(\Psi,\Theta, \Phi)$, hence
\begin{equation}
  \label{S3}
  G_{AB} dX^A dX^B = d\Psi^2 + \sin^2 \Psi \left( d\Theta^2 + \sin^2 \Theta\: d\Phi^2\right).
\end{equation}
In addition, we assume that the map $U$ is spherically $\ell$-equivariant, that is
\begin{equation}
  \label{eq:WM.ansatz}
  \Psi = u(t,r),\qquad (\Theta,\phi)=\Omega_{\ell}(\theta,\phi)\,,
\end{equation}
where $\Omega_{\ell}:S^2\mapsto S^2$ is a harmonic eigenmap with eigenvalue $\ell(\ell+1)$ ($\ell\in \mathbb{N}$).
Under these assumptions the wave map equation
\eqref{WM} reduces to the single scalar wave equation
\begin{equation}
  \label{WMrad}
  -\ddot{u} +u'' + \frac{2}{r}u' - \frac{\llone}{2r^2}\sin 2u=0\,,
\end{equation}
where dots and primes  denote derivatives with respect to $t$ and $r$, respectively.

Eq.\eqref{WMrad}
 is invariant under scaling: if $u(t,r)$ is a solution, so is
$u_{\lambda}(t,r)=u(t/\lambda, r/\lambda)$.
 The total conserved energy scales as $E[u_{\lambda}] =
\lambda E[u]$, which means Eq.\eqref{WMrad} is supercritical and singularities are expected to develop for some initial data. In fact, it is known that Eq.\eqref{WMrad} has infinitely many self-similar solutions \cite{b1} which are explicit examples  of singularities forming in finite time from regular initial data \cite{shatah}. Moreover, the ground state self-similar solution is stable and acts as an attractor for generic large initial data \cite{bct,don}.  Note also that the scaling symmetry precludes existence of nontrivial static solutions. The formation of singularities for generic initial data and the absence of nontrivial static solutions make Eq.\eqref{WMrad} uninteresting from a physical point of view. One way to cure these deficiencies  is to modify the action \eqref{action} by adding a suitable higher order term with subcritical scaling. Such a term breaks the scaling symmetry and prevents shrinking of solutions to zero size. The best known modification of this type is the Skyrme model \cite{skyrme}. This model admits stable static solutions (so called skyrmions) and enjoys global regularity in time \cite{li}. Unfortunately, these nice features are achieved at the expense of a significant increase of complexity: the evolution equation is quasilinear and very difficult to analyze (note that it took fifty years to prove that no singularities can form during the evolution). The numerical evidence for the soliton resolution conjecture in the Skyrme model was given in \cite{bcr} but to our knowledge there is no corresponding rigorous result.

Motivated by a desire to have a simpler model than Skyrme's and yet possessing similar properties
 we proceed in a different manner. We keep the action unchanged but modify the domain of the wave map. Namely, we consider Eq.\eqref{WMrad} outside a ball of radius $R$, that is for $r\geq R$
and impose the Dirichlet condition at the boundary $u(t,R) =
0$ \cite{bss}.
 Hereafter, we choose the unit of length such that $R=1$. We are interested in the long-time behavior of solutions starting from smooth initial data with finite energy
\begin{equation}
  \label{energy}
  E(u):= \frac{1}{2}\int_{1}^{\infty}
  \left( r^2 \dot{u}^2 + r^2 u'^2 +\ell(\ell+1)\sin^2{\!u} \right) \ud r < \infty\,.
\end{equation}
Note that  finite energy solutions must satisfy the boundary condition at infinity $u(t,\infty)=N\pi$, where an integer $N$ is the topological degree of the map
(sometimes called the winding
number). Since  the degree cannot change during the evolution, the initial value problem for Eq.\eqref{WMrad} breaks into infinitely many disjoint topological sectors labeled by the degree $N$. Note also that global regularity in time is automatically guaranteed because by spherical symmetry singularities for Eq.\eqref{WMrad} cannot occur for $r>0$. In other words, by cutting off the ball from the domain of the map we make the equation effectively subcritical.

 Knowing that solutions exist for all times, it is natural to ask  how they behave as $t\rightarrow \infty$. Below we show that for each $\ell\in \mathbb{N}$ and each $N\in \mathbb{Z}$ there exist a unique static solution $u_{\ell,N}(r)$ which is linearly stable. We conjecture that the solution $u_{\ell,N}$ is a universal global attractor, that is all solutions starting with smooth initial data of degree $N$ asymptotically converge to it. The numerical evidence supporting this conjecture is presented in section~5.
\section{Static solutions}
To study static solutions it is convenient to use  new variables $x=\ln{r}$ and $v(x)=u(r)$.  Then, time-independent Eq.\eqref{WMrad} takes the autonomous form
\begin{equation}
  \label{static.x}
  v''+v' - \frac{\llone}{2}\sin 2v = 0\,,
\end{equation}
which in mechanics describes the motion of a pendulum with constant friction.
A regular finite energy solution of degree $N$ corresponds to a trajectory which in the phase plane $(v,v')$ starts from a point $(0,a)$ and goes to the point $(N\pi,0)$ as $x\rightarrow \infty$. The existence and uniqueness of such a trajectory for each  $N$ follows from elementary phase-plane analysis. The phase portrait displaying the flow of the vector field $X=(v',-v'+\frac{1}{2}\ell(\ell+1)\sin{2v})$ is depicted in Fig.~1. The flow has stable focal points at $\left((N+1/2)\pi,0)\right)$ with eigenvalues $\alpha=(-1\pm i\sqrt{4\ell(\ell+1)-1})/2$ and saddle points at $(N\pi,0)$ with eigenvalues $\alpha=\ell$ and $\alpha=-\ell-1$. From each saddle point there emerge two separatrices corresponding to the stable ($\alpha=-\ell-1$) and unstable ($\alpha=\ell$) manifolds. It is routine to show that close to the saddle the stable manifold can be parametrized in the form
\begin{equation}
  \label{stable_manifold}
  v(x) = N\pi-b\, e^{-\lone x}+ \mathcal{O}\left(e^{-3\lone x}\right)\,,
\end{equation}
where the coefficients of all higher order terms are determined by the parameter $b$.
Generically the point $(0,a)$ belongs to the  basin of attraction of one of the focal points but for isolated values $a=a_{\ell,N}$ this point lies on the separatrix which runs into the saddle $(N\pi,0)$ along the stable manifold.
 %===============================..figure..=====================================%
\begin{figure}[h]
  \begin{center}
    \includegraphics[width=0.5\textwidth]{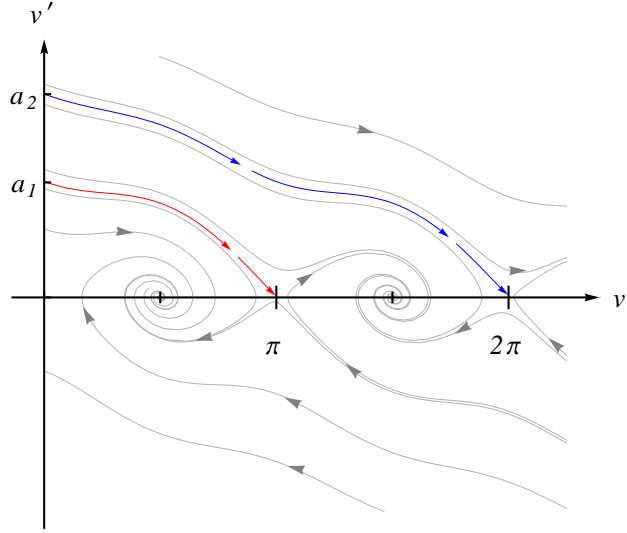}
    \caption{The phase portrait for Eq.\eqref{static.x} for $\ell=2$.}
    \label{fig:static.profiles.all}
  \end{center}
\end{figure}
% ===============================..figure..=====================================%

Translating the above analysis back to the original variables we conclude that for each integer $N$ there exists a unique static solution $u_{\ell,N}(r)$ of Eq.\eqref{WMrad} satisfying the boundary conditions $u_{\ell,N}(1)=0$ and $u_{\ell,N}(\infty)=N\pi$. This solution is parametrized by the  pair of numbers $(a_{\ell,N},b_{\ell,N})$ such that $u'_{\ell,N}(1)=a_{\ell,N}$ and
\begin{equation}\label{uasym}
u_{\ell,N}(r)=N\pi -\frac{b_{\ell,N}}{r^{\ell+1}} +\mathcal{O}(r^{-3\lone})\quad \mbox{as} \quad r\rightarrow\infty\,.
\end{equation}
The parameters $a_{\ell,N}$ and $b_{\ell,N}$  can be determined numerically with the help of a shooting method. A few values of these parameters for different $\ell$ and $N$ are listed in
Table~I, while
Fig.~2 depicts several profiles $u_{\ell,N}(r)$. Let us note that the static solutions described above were first found by Balakrishna et al. \cite{bss} who used them as a simple model of nucleons.

\begin{table}[!h]
  \centering
  \begin{tabular}{|c|cccccc|}\hline
    \noalign{\smallskip}
     $\,\,\,\,$ & $\ell=1$ & $\ell=2$ & $\ell=3$ & $\ell=4$ & $\ell=5$ & $\ell=6$\\
    \noalign{\smallskip}\hline\noalign{\smallskip}
    $\,\,a_{\ell,1}\,\,$ & 3.786299 & 4.327397 & 4.800690 & 5.226581 & 5.617059 & 5.979801 \\
    $\,\,b_{\ell,1}\,\,$ & 4.847841 & 6.147165 & 7.240761 & 8.203436  & 9.073200  & 9.872569 \\[1ex]\hline\hline
    $\,\,a_{\ell,2}\,\,$ & 6.983263 & 7.601218 & 8.161078 & 8.677933 & 9.161071  & 9.616701 \\
    $\,\,b_{\ell,2}\,\,$ &  15.810784  & 29.162429  & 46.410705 & 67.463775 & 92.172786 & 120.372175 \\[1ex]
    \hline
  \end{tabular}
  \caption{Parameters of the static solutions.}
  \label{tab:static.params}
\end{table}
% ===============================..table..=====================================%
\noindent \emph{Remark~1.} It does not seem feasible to express the solutions $u_{\ell,N}(r)$ in closed form, however  for large $\ell$ one can obtain a good analytic approximation for $u_{\ell,1}$ as follows.
 Let~$y=(\ell+1) x$ and $\bar v(y)=v(x)$. Then, in the limit $\ell\rightarrow\infty$ Eq.\eqref{static.x} becomes
$
  \bar v''- \frac{1}{2}\sin 2\bar v = 0
$.
The solution of this limiting equation satisfying the boundary condition $\bar v(\infty)=\pi$ is
$\bar v(y)=2 \arctan\left(A e^y\right)$, where $A$ is an integration constant.  Returning to the variable $x$ and setting $A=2/b_{\ell,1}$ (to agree with \eqref{stable_manifold}), we obtain the large $\ell$ approximation
$v(x)\approx 2 \arctan\left(\frac{1}{2b_{\ell,1}} e^{(\ell+1)x}\right)$, which is valid outside the boundary layer  of width $\mathcal{O}(\ell^{-1})$ near $x=0$.  Using matched asymptotics one can systematically match this outer solution to the inner solution satisfying the boundary condition $v(0)=0,v(0)=a_{\ell,1}$, however for practical purposes  the following ad hoc modification (which does not alter the large $x$ asymptotics) is good enough
\begin{equation}\label{v1approx}
v_{\ell,1}(x)\approx \bar v_{\ell,1}(x)=2\arctan\left(\frac{2}{b_{\ell,1}} (e^{(\ell+1)x} -e^{-(\ell+2) x}) \right)\,.
\end{equation}
 As shown in Fig.~2, this  approximation  is quite accurate already for $\ell=1$.
 \vskip 0.3cm
% ===============================..figure..=====================================%
\begin{figure}[h]
  \begin{center}
    \includegraphics[width=0.48\textwidth]{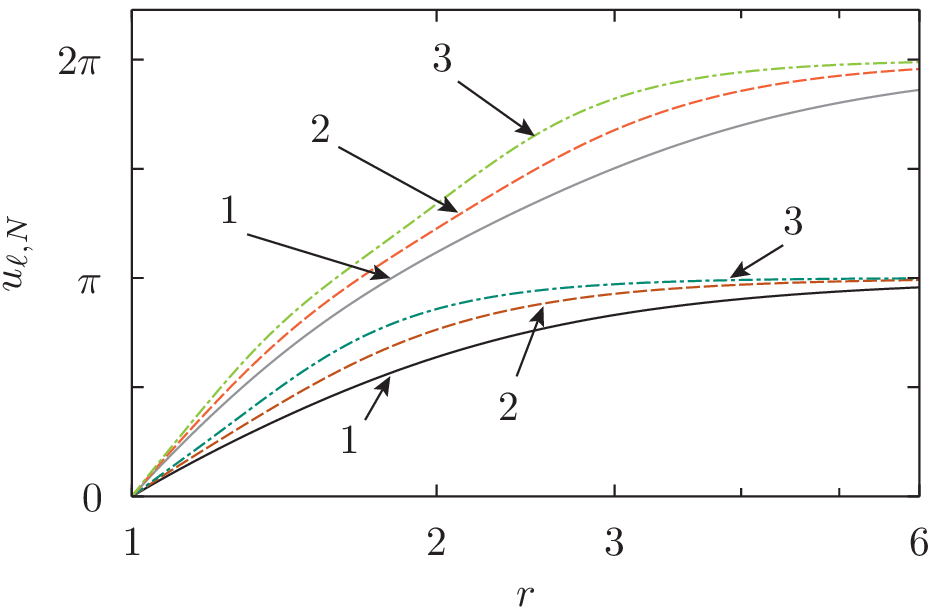}
    \includegraphics[width=0.48\textwidth]{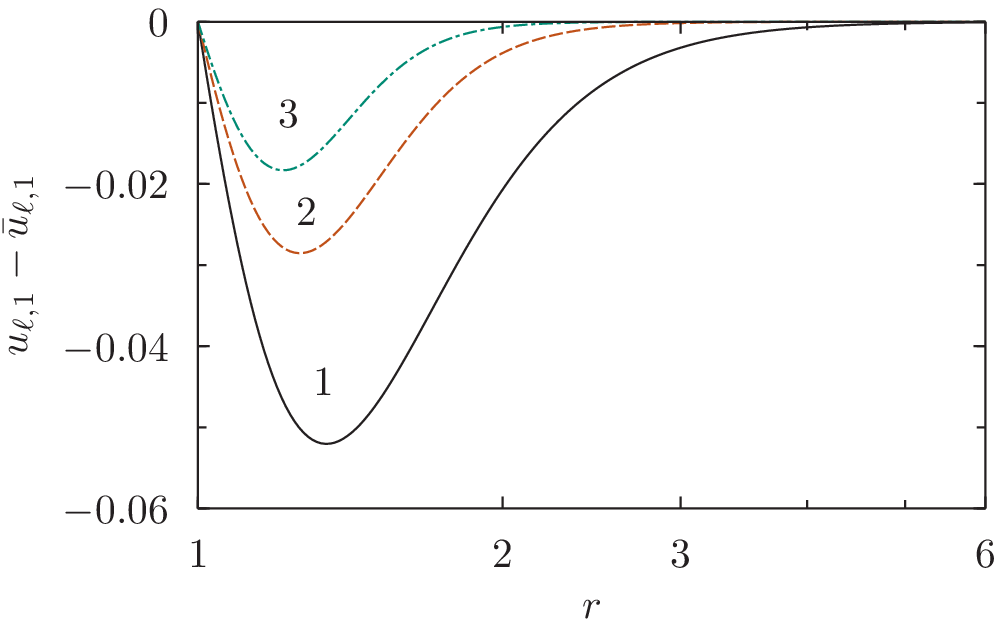}
    \caption{Left: Profiles of the static solutions for $N=1,2$ and $\ell=1,2,3$. Right: Deviation of the approximate solution $\bar u_{\ell,1}(r)=\bar v_{\ell,1}(\ln{r})$ from the true solution $u_{\ell,1}(r)$.}
    \label{fig:static.profiles.all}
  \end{center}
\end{figure}
% ===============================..figure..=====================================%
\vspace{-1cm}
\section{Linearization about static solutions}

Substituting $u(t,r)=u_{\ell,N}(r)+w(t,r)$ into Eq.\eqref{WMrad} and linearizing we obtain the evolution equation for small perturbations around the static solution
\begin{equation}\label{eqw}
 \ddot{w} -w'' - \frac{2}{r}w' + \frac{\ell(\ell+1)}{r^2}\,w+ V_{\ell,N}(r) w=0\,,\qquad  V_{\ell,N}(r) = -\frac{2\llone}{r^2} \sin^2{u_{\ell,N}(r)}\,.
\end{equation}
Assuming harmonic time dependence, $w(t,r)=e^{-i\lambda t} \psi(r)$, we get the eigenvalue problem
\begin{equation}
  \label{eqpsi}
  -\psi''-\frac{2}{r} \psi'+\frac{\ell(\ell+1)}{r^2}\,\psi+V_{\ell,N}(r) \psi = \lambda^2 \psi
\end{equation}
for $\psi\in L_2\left([1,\infty),r^2 dr\right)$ and satisfying $\psi(1)=0$.
The spectrum is purely continuous $\lambda^2\geq 0$. To see this, note that $\psi(r)=r \frac{du_{\ell,N}(r)}{dr}$ satisfies Eq.\eqref{eqpsi} for $\lambda^2=0$, as can be readily checked by differentating the time-independent Eq.\eqref{WMrad} with respect to $r$. The presence of this zero mode  is the consequence of scale invariance of Eq.\eqref{WMrad}. The zero mode  is not an eigenfunction because it does not vanish at $r=1$, nonetheless the fact that it has no zeros (since $u_{\ell,N}(r)$ is monotone) implies by the Sturm oscillation theorem that there are no negative eigenvalues. Therefore, the static solutions $u_{\ell,N}$ are linearly stable.
% ===============================..figure..=====================================%
\begin{figure}[!h]
  \begin{center}
    \includegraphics[width=0.48\textwidth]{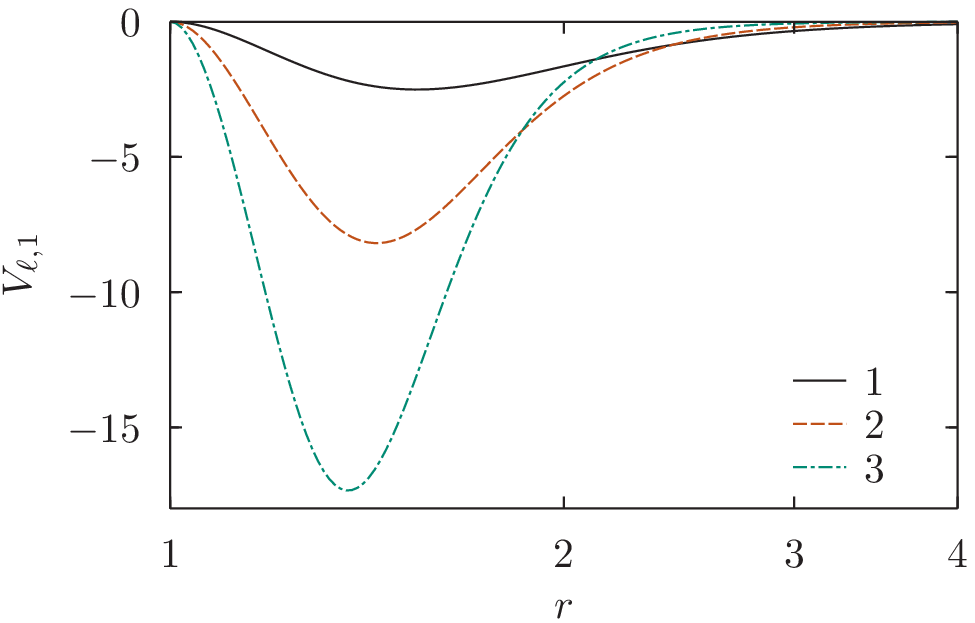}
     \includegraphics[width=0.48\textwidth]{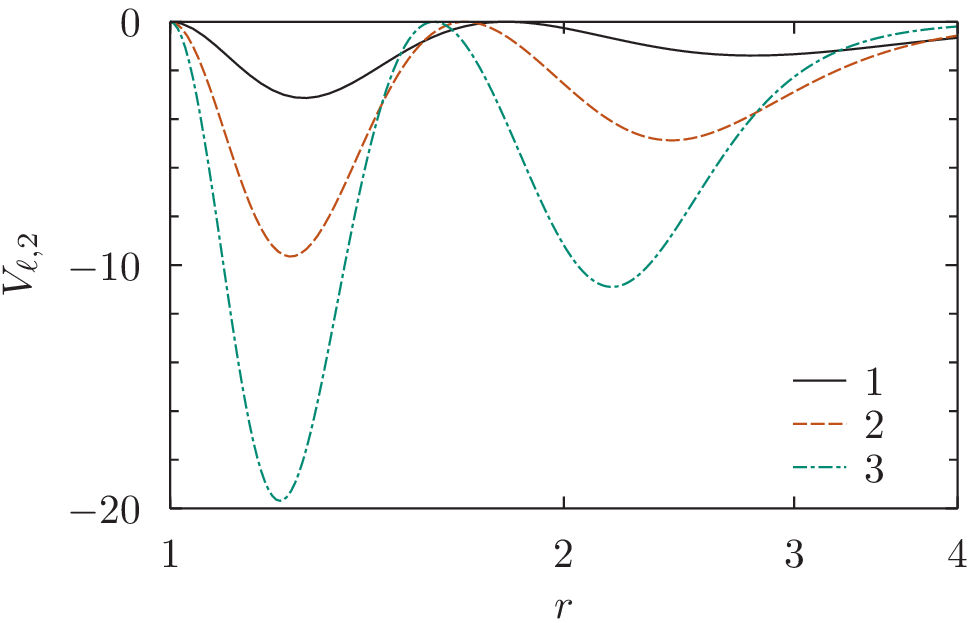}
    \caption{The potentials $V_{\ell,N}(r)$  for $N=1,2$ and $\ell =1, 2, 3$.}
    \label{fig:efpot.n1}
  \end{center}
\end{figure}
% ===============================..figure..=====================================%

The linear stability is only the first necessary step in establishing asymptotic stability. To understand the latter we need to determine the quasinormal modes (also called complex resonances or scattering frequencies), that is solutions of Eq.\eqref{eqpsi} which vanish at $r=1$ and satisfy the outgoing wave condition $\psi(r) \sim \frac{1}{r}e^{i\lambda r}$ with $\Im(\lambda)<0$  for  $r\rightarrow \infty$.
The fundamental (that is, the least damped) quasinormal mode is expected to govern an intermediate phase of the relaxation to the static solution.

 Let us first describe the quasinormal modes about the  trivial solution $u_{\ell,0}=0$. In this case Eq.\eqref{eqpsi}
takes the form of the spherical Bessel equation
\begin{equation}
  \label{eqpsi0}
  -\psi''-\frac{2}{r}\psi'+ \frac{\llone}{r^2} \psi = \lambda^2 \psi\,,
\end{equation}
 whose exact outgoing solutions are given by the spherical Hankel functions of the first kind  $\psi(r)=h_{\ell}^{(1)}(\lambda r)$. Thus, the quantization condition for the eigenfrequencies of quasinormal modes is
 $\psi(1)=h_{\ell}^{(1)}(\lambda)=0$ which is a polynomial equation of order $\ell$, hence has exactly $\ell$ roots. For example, using the formulae for the first two spherical Hankel functions
 \begin{equation}\label{hankel}
 h_1^{(1)}(z)=e^{i z} \left(\frac{1}{i z^2}-\frac{1}{z}\right)\,,\qquad
 h_2^{(1)}(z)=e^{i z} \left(\frac{3}{i z^3}-\frac{3}{z^2}-\frac{1}{i z}\right)\,,
 \end{equation}
 one gets $\lambda=-i$ for $\ell=1$ and $\lambda=\pm \sqrt{3}/2-3i/2$ for $\ell=2$. From the known results on zeros of Hankel's functions  it follows that the eigenfrequency of the fundamental quasinormal mode for large $\ell$  is  (see Proposition E.2 in \cite{sw})
 \begin{equation}\label{hankel_zero}
 \lambda_{\ell} = \nu - \zeta \nu^{\frac{1}{3}} +\frac{3}{10} \zeta^2 \nu^{-\frac{1}{3}} +\mathcal{O}\left(\nu^{-1}\right)\,,
  \end{equation}
  where $\nu=\ell+1/2$, $\zeta=2^{-\frac{1}{3}} e^{-\frac{2\pi i}{3}}\eta_1$, and $\eta_1\simeq -2.338$ is the first zero of the Airy function.

 For $N\neq 0$ it is not possible to calculate the quasinormal modes analytically, e.g. because the potential $V_{\ell,N}(r)$ in \eqref{eqpsi} is not known in closed form, thus we must resort to numerical methods.   We use a shooting method which goes as follows. First, we introduce  new independent and dependent variables
   \begin{equation}\label{newvar}
   \rho=\frac{1}{r}\,,\quad \eta(\rho)=r e^{-i\lambda r} \psi(r)\,.
    \end{equation}
   The advantage of this transformation is two-fold: i) it peels off the oscillatory behavior of $\psi(r)$ at infinity and ii)
     compactifies the semi-infinite interval $1\leq r<\infty$ to the finite interval $0\leq \rho \leq 1$ (switching the endpoints). In terms of the new variables Eq.\eqref{eqpsi} becomes
   \begin{equation}\label{eqeta}
   \rho^2 \eta''+(2\rho-2 i \lambda) \eta' -\ell(\ell+1)\eta +2\ell(\ell+1)\sin^2\left(u_{\ell,N}\right)\eta=0\,.
   \end{equation}
   Next, we solve this equation numerically away from the endpoints starting from the desired "initial values". At $\rho=1$ the initial value is  $\eta(1)=0$ (and an arbitrary nonzero $\eta'(1)$). At $\rho=0$ the desired  outgoing solution is defined by the asymptotic expansion
   \begin{equation}\label{series0}
    \eta(\rho) \sim \sum_{j=0}^{\ell} \frac{i^j (\ell+j)!}{(2\lambda)^j j! (\ell-j)!}\, \rho^j
    +\sum_{j=2\ell+3}^{\infty} c_j \rho^j\,.
   \end{equation}
   The first finite sum in \eqref{series0} corresponds to the free outgoing solution. The second infinite sum is the divergent asymptotic series (since $\rho=0$ is the irregular singular point) whose coefficients $c_j(\lambda,b_{\ell,N})$ can be calculated recursively using \eqref{uasym}. We use the series \eqref{series0} to impose the initial value at  $\rho=0.1$. Note that the unwanted ingoing solution is of the order $\mathcal{O}(e^{-\frac{2|\Im(\lambda)|}{\rho}})$, hence it is difficult  to keep track of it numerically, especially if $|\Im(\lambda)|$ is large. To suppress the contamination of the initial value at $\rho=0.1$ by this exponentially small ingoing admixture it is crucial to compute the series \eqref{series0} with very high precision. We achieve this precision by the optimal truncation of the asymptotic series (alternatively, the Pad\'e summation could be used). The requirement that the logarithmic derivatives of the two numerical solutions launched from $\rho=0.1$ and $\rho=1$ join smoothly at a midpoint, say $\rho=0.5$, serves as the quantization condition
   for quasinormal modes.  The fundamental eigenfrequencies determined in this manner are listed in
Table~II. We denote the $j$-th ($j=0,1,\dots$) quasinormal eigenfrequency and eigenmode by $\lambda_{\ell,N}^j=\Omega_{\ell,N}^j-i\, \Gamma_{\ell,N}^j$ and $\psi_{\ell,N}^j(r)$.
   We adopt the convention that for given $N$ and $\ell$ the damping rate $\Gamma_{\ell,N}^j$ increases with $j$, hence the $j=0$ mode  is fundamental. Below, to avoid notational clutter, we drop the index $j=0$ on the fundamental quasinormal modes.

 Note that as $\ell$ increases, the frequencies $\Omega_{\ell,N}$ increase  while the damping rates $\Gamma_{\ell,N}$ rapidly decrease to zero. This behavior was intuitively to be expected  in view of the fact that the potential $V_{\ell,N}(r)$ becomes narrower and deeper as  $\ell$ grows (see Fig.~3) (hence the waves get increasingly  squeezed and trapped as $\ell$ grows).
% ===============================..table..=====================================%
\begin{table}[!h]
  \centering
  \begin{tabular}{|cc|cccccc|}\hline
    \noalign{\smallskip}
     & $\,\,\,\,\,$ & $\ell=1$ & $\ell=2$ & $\ell=3$ & $\ell=4$ & $\ell=5$ & $\ell=6$ \\
    \noalign{\smallskip}\hline\noalign{\smallskip}
   & $\,\,\Omega_{\ell,1}\,\,$ & $\,$ 0.425517 $\,$ & $\,$ 0.910515$\,$ &$\,$ 1.346222 $\,$ & $\,$1.752249$\,$ & $\,$2.129364$\,$ & $\,$ 2.475564 $\,$  \\
   & $\,\,\Gamma_{\ell,1}\,\,$ & $\, 0.347445 \,$ &$\, 0.246627 \,$& $\,0.151154\,$
   & $\,0.080210 \,$ & $\,0.035763 \,$ & $\, 0.012945 \,$
    \\[1ex]\hline\hline
   & $\,\,\Omega_{\ell,2}\,\,$ & $\,0.228006\,$ & $\,0.436739\,$ & $\, 0.632007 \,$ & $\,0.814959\,$
   & $ \, 0.982609 \,$ & $\, 1.137494 \,$  \\
   & $\,\,\Gamma_{\ell,2}\,\,$ & $\, 0.121276 \,$ & $\,0.058779 \,$ & $\,0.021354\,$ & $\,\SI{5.2862e-3}\,$
   & $\,\SI{8.6491e-4}\,$ & $\, \SI{1.0037e-4} \,$
     \\[1ex]
    \hline
  \end{tabular}
  \caption{The fundamental eigenfrequencies  $\lambda_{\ell,N}=\Omega_{\ell,N}-i\, \Gamma_{\ell,N}$ for $N=1,2$.}
  \label{qnm}
\end{table}
% ===============================..table..=====================================%

% ===============================..figure..=====================================%
\begin{figure}[!h]
  \begin{center}
    \includegraphics[width=0.6\textwidth]{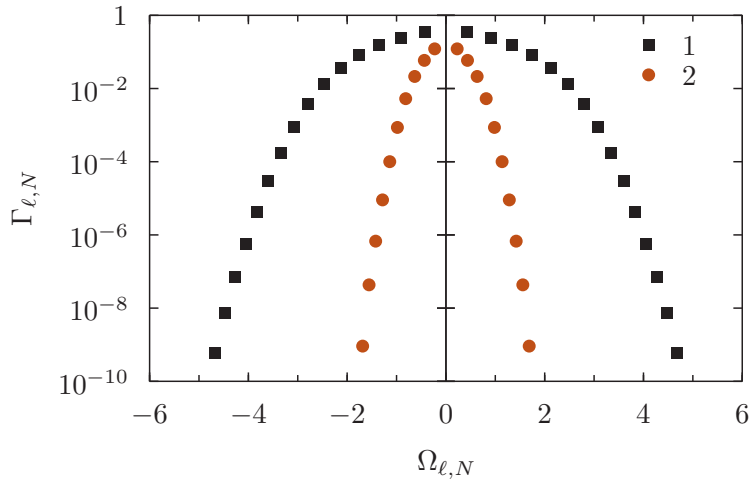}
    \caption{The distribution of fundamental eigenfrequencies  $\lambda_{\ell,N}$ for $N=1,2$ and $\ell=1,2,\dots,15$ in the complex $\lambda$--plane. Numerics indicates that  $\Omega_{\ell,N}\sim \ell^{1/2}$ and $\Gamma_{\ell,N} \sim e^{-\ell(\ell+1)}$ for large $\ell$.}
    \label{fig:efpot.n1}
  \end{center}
\end{figure}
% ===============================..figure..=====================================%

To reassure ourselves and verify independently the above computation we solved numerically the linearized evolution equation \eqref{eqw}. According to the scattering  theory, for intermediate times the solution at a fixed point is space can be represented by the sum of quasinormal modes
\begin{equation}\label{qnm_exp}
    w(t,r) \sim \sum_{j} A_j \exp\left(-i\lambda_{\ell,N}^j t\right) \psi_{\ell,N}^j(r)\,,
\end{equation}
where the coefficients $A_j$ depend on initial data. After a short transient time, all the quasinormal modes but  the fundamental one die out (see Fig.~5), hence we can readily determine the fundamental eigenfrequency by fitting the exponentially damped sinusoid to the numerical solution. The results of such a fit agree to four (for $\ell=1,2$) and six  (for $\ell\geq 3$) decimal places with the values given in Table~II. To compute the parameters of excited modes we applied the Prony method \cite{prony}. The outcome of this computation suggests that there are exactly $N$ quasinormal modes about the solution $u_{\ell,N}$.
\vskip 0.5cm
% ===============================..figure..=====================================%
\begin{figure}[!h]
  \begin{center}
    \includegraphics[width=1.0\textwidth]{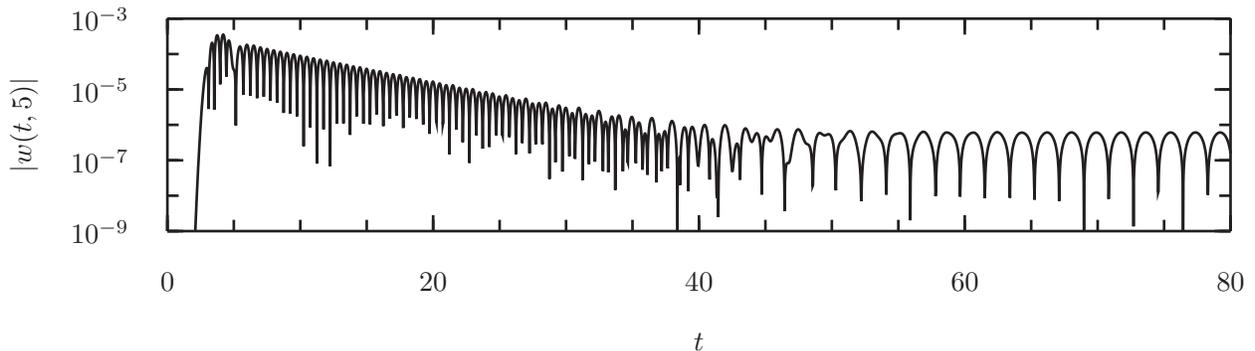}
    \caption{Quasinormal ringdown to zero for the solution of the linearized equation \eqref{eqw} for $N=2$ and $\ell=10$. After the passage of the direct signal from the initial data, an observer sitting at $r=5$ sees two clearly pronounced  trains of exponentially damped oscillations which correspond to the first excited quasinormal mode with $\lambda_{10,2}^1=6.2925-i\, 0.1718$ and the fundamental quasinormal mode with $\lambda_{10,2}=1.6862-i\, 9.1606\times 10^{-10}$, respectively.}
    \label{fig:efpot.n1}
  \end{center}
\end{figure}
% ===============================..figure..=====================================%
\noindent\emph{Remark~2.}  For very late times the linear evolution described by Eq.\eqref{eqw} changes character from the quasinormal ringdown (which becomes negligible) to the polynomial decay called a tail.
The tail is due to the backscattering of waves  at large distances and its decay rate is insensitive to the presence of the ball. According to the linear scattering theory (see e.g. \cite{ching}), for the potential falling off as $r^{-\beta}$ ($\beta>3$) for large $r$ and compactly supported initial data, the tail decays as $t^{-\gamma}$, where $\gamma=2\ell+\beta$. In the case at hand, it follows from \eqref{uasym} and \eqref{eqw} that $V_{\ell,N}(r)\sim r^{-(2\ell+4)}$ as $r\rightarrow\infty$, hence $\gamma=4\ell+4$. We note for completeness that for the initial data which are not compactly supported and fall off polynomially at spatial infinity there is also a direct polynomially decaying in time signal from the data.
\section{Numerical evidence}
In this section we solve numerically Eq.\eqref{WMrad} with the Dirichlet boundary condition $u(t,1)=0$ and
 smooth finite energy initial data of various topological degrees $N$.  The numerical technique is standard. We use the method of lines with
a fourth-order Runge-Kutta time integration and fourth-order spatial finite differences. Our computational domain is large enough so that its outer boundary is causally disconnected from the region of interest. To ensure numerical stability, the high frequency modes were damped using the Kreiss-Oliger dissipation. To suppress round-off errors, all
simulations were performed in the quadruple (128-bit) arithmetic precision.
It is worth stressing  that the Dirichlet boundary condition  and smoothness  imply that the Taylor series expansion at $r=1$ takes the form
\begin{equation}\label{reg}
    u(t,r)=a(t) (r-1) - a(t) (r-1)^2 +\mathcal{O}\left((r-1)^3\right)\,,
\end{equation}
where a free function $a(t)$ determines recursively all the coefficients of the series. Taken at $t=0$, the expansion \eqref{reg} and its time derivative express the compatibility conditions between the initial values and the Dirichlet boundary condition.
 % ===============================..figure..=====================================%
\begin{figure}[!h]
  \begin{center}
    \includegraphics[width=1.0\textwidth]{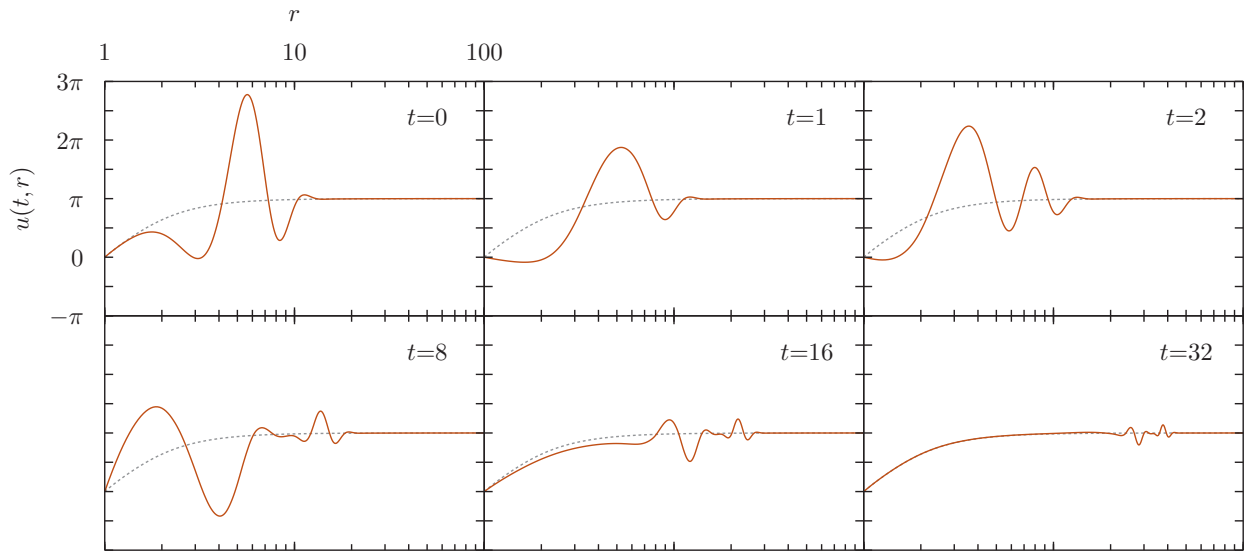}
    \caption{
    Convergence to the static solution $u_{1,1}(r)$ (dashed line).}
    \label{fig:efpot.n1}
  \end{center}
\end{figure}
% ===============================..figure..=====================================%

 Our main observation (illustrated in Fig.~6) is that for all $\ell$ every solution of Eq.\eqref{WMrad} satisfying $u(t,1)=0$ and starting from smooth finite energy initial data of degree $N$ converges (in any compact interval) to the static solution $u_{\ell,N}$ as $t\rightarrow \infty$. A rigorous proof (and a precise formulation) of this claim is left as a challenge to the PDE experts.

Below we present some more quantitative aspects  of the relaxation process.
We focus on pointwise convergence in the timelike direction, that is we look at the time development of $u(t,r)$ at a fixed point in space. We consider two kinds of initial data: (i) spatially localized perturbations of  static solutions and (ii) arbitrary initial data with different next-to-leading order asymptotic behavior at spatial infinity than static solutions. We illustrate the case (i) with initial data of the form of the "kicked" static solution
\begin{equation}\label{fig7}
    u(0,r)=u_{\ell,N}(r),\quad  \dot u(0,r)=\varepsilon (r-1)^3 \exp\left(-\frac{(r-r_0)^2}{2\sigma^2}\right)
\end{equation}
with parameters $\varepsilon=0.8,r_0=1.5, \sigma=0.1$,
and the case (ii) with initial data
\begin{equation}\label{fig8}
    u(0,r)=\frac{2(r-1)^2}{1+(r-1)^2}\, \arctan(r-1), \quad \dot u(0,r)=0\,.
\end{equation}
In both cases the intermediate phases of evolution are similar and have the form of quasinormal ringdown, which is dominated by the fundamental quasinormal mode. Thus, this phase is well described by the linearized approximation
(see Figs.~7,8,9).
\vspace{0.3cm}
% ===============================..figure..=====================================%
\begin{figure}[!h]
  \begin{center}
    \includegraphics[width=0.49\textwidth]{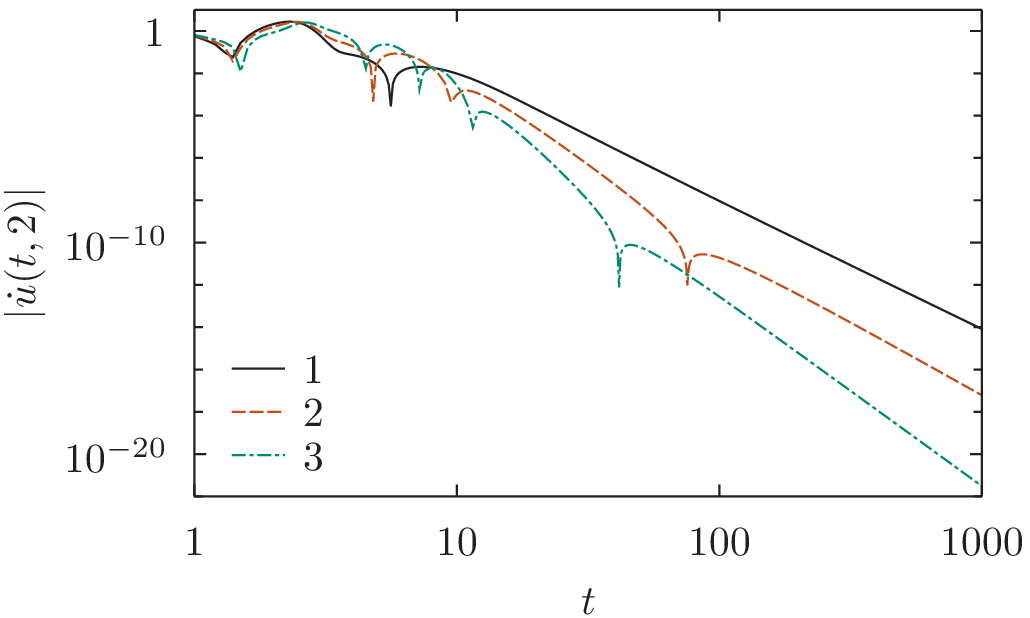}
    \includegraphics[width=0.49\textwidth]{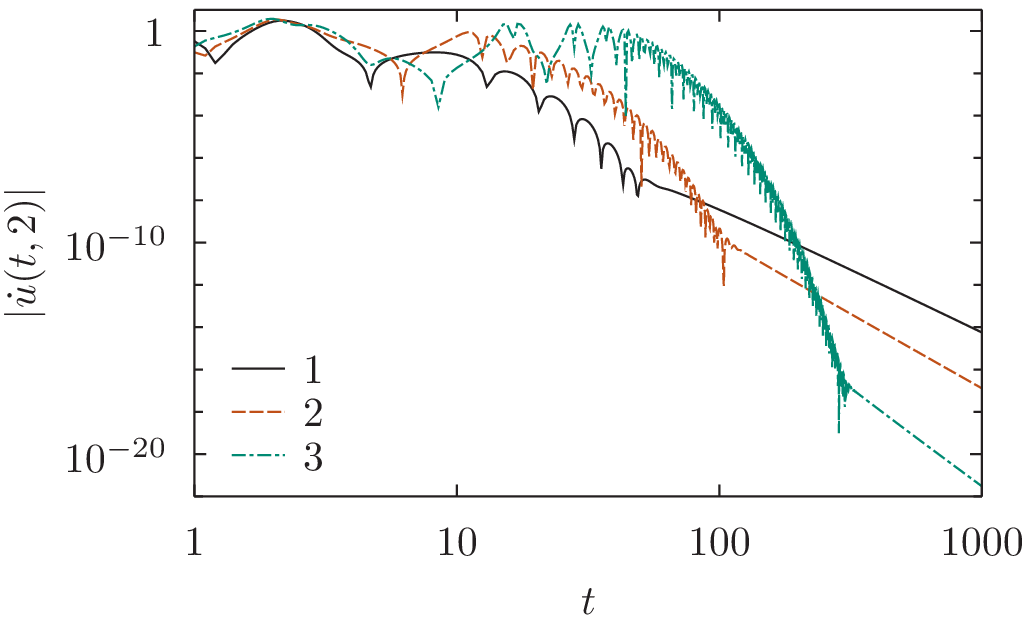}
    \caption{Pointwise convergence to the static solution for $\ell=1,2,3$ and $N=0$ (left panel) and $N=1$ (right panel) for exponentially localized perturbations of the static solution of the form \eqref{fig7}.
    Quasinormal ringdown for intermediate times  yields to a polynomial tail for late times. In both cases the tail decays as $t^{-5}$ (for $\ell=1$) and as $t^{-(2\ell+2)}$ (for $\ell\geq 2$).  Note that for $N\geq 1$, as $\ell$
    increases, the duration of the ringdown phase rapidly increases because the damping rate of the fundamental quasinormal mode decreases while the decay rate of the tail increases.}
    \label{fig:efpot.n1}
  \end{center}
\end{figure}

% ===============================..figure..=======================
\begin{figure}[!h]
  \begin{center}
    \includegraphics[width=1.0\textwidth]{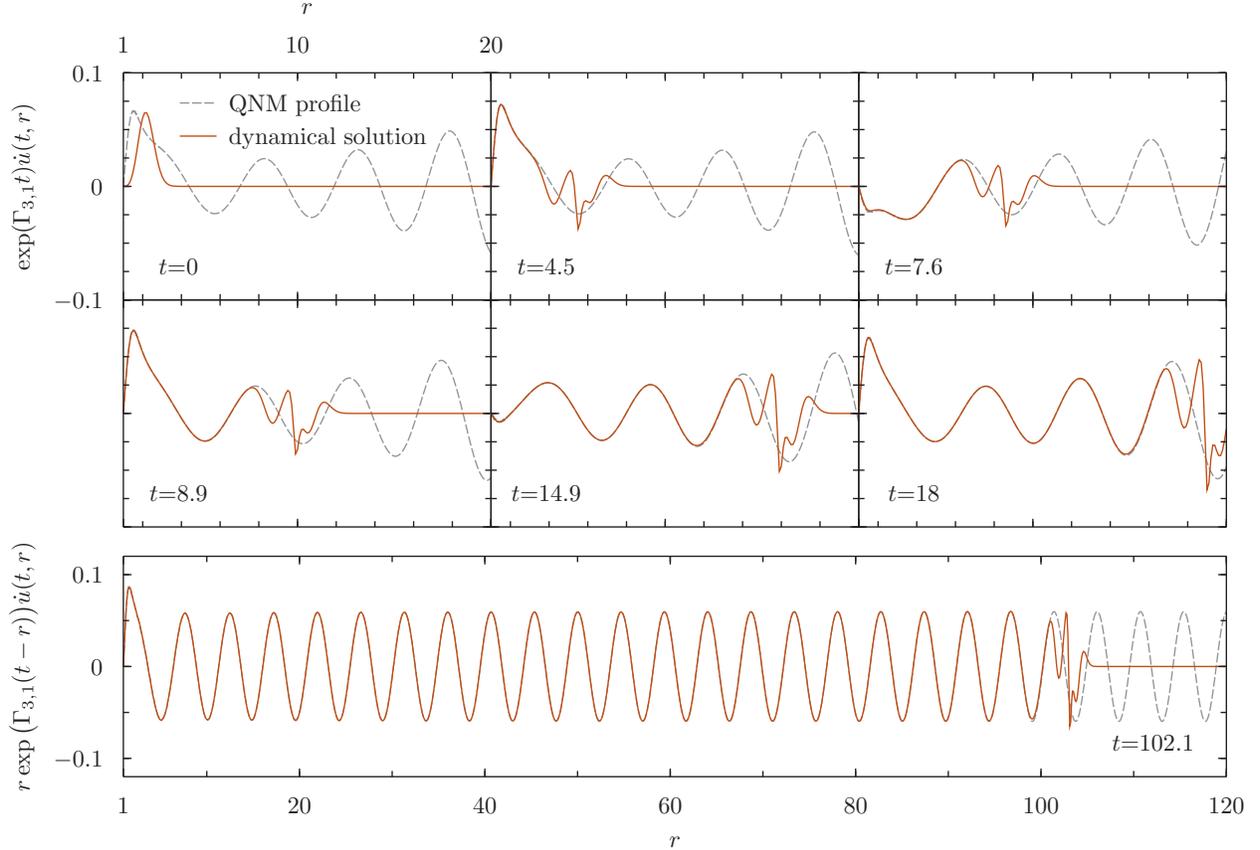}
    \caption{Snapshots from the evolution starting from initial data \eqref{fig7} for $\ell=3,N=1$.  Behind the wavefront, the spatial profile of the solution approaches  the spatial profile of the fundamental quasinormal eigenfunction $A\,\Re\left(e^{-i \lambda_{3,1}(t-t_0)}\psi_{3,1}(r)\right)$. The parameters $A$ and $t_0$ were fitted at $t=4.5$ once for all times.  For better visibility the profiles are multiplied by  $e^{\Gamma_{3,1} t}$ (upper plots) and  $r e^{\Gamma_{3,1} (t-r)}$ (bottom plot). }
    \label{fig:efpot.n1}
  \end{center}
\end{figure}

The late time behavior is more complicated because there are three competing contributions to the tail: two kinds of linear tails mentioned in Remark~2 and the nonlinear tail caused by self-interaction of the field. The latter was described in detail in  \cite{BCRZ} in the framework of weakly nonlinear perturbation analysis, in particular it was shown there that the nonlinear tail
decays as $t^{-5}$ for $\ell=1$ and as $t^{-(2\ell+2)}$ for $\ell\geq 2$. For the initial data \eqref{fig7} this tail dominates the late time evolution (Fig.~7) while for the initial data \eqref{fig8} the linear tail coming from the deviation of asymptotic behavior of the data from the static solution is dominating (Fig.~9).
% ===============================..figure..=====================================%
\begin{figure}[!h]
  \begin{center}
    \includegraphics[width=0.6\textwidth]{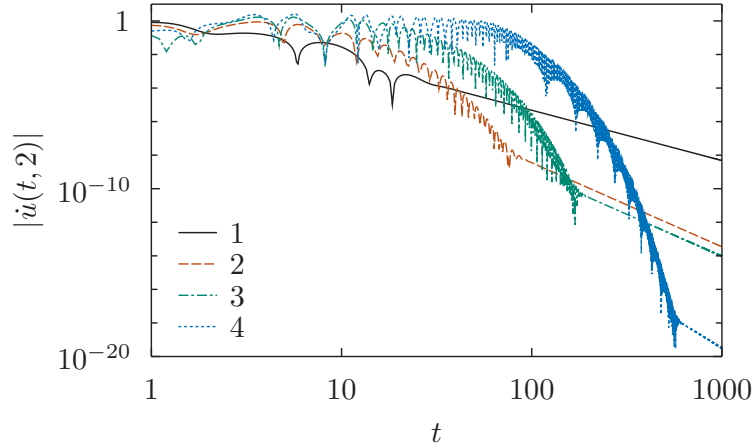}
    \caption{Convergence to the static solution
    $u_{\ell,N}(r)$ for $\ell=1,2,3,4$ and $N=1$ for initial data \eqref{fig8}. The deviation of these data from  the static solution is of the order $\mathcal{O}(r^{-(\ell+1)})$ for large $r$ which leads to the linear tail decaying as $t^{-\gamma}$ with $\gamma=2$ ($\ell=1$), $\gamma=4$ ($\ell=2$), $\gamma=4$ ($\ell=3$), and  $\gamma=6$ ($\ell=4$). The nolinear tail  is subdominant.}
    \label{fig:efpot.n1}
  \end{center}
\end{figure}
% ===============================..figure..=====================================%

\subsection{Acknowledgments}
We thank Patryk Mach for helpful remarks.
This work was supported in part by the NCN grant NN202 030740.


\begin{thebibliography}{10}
\bibitem{bss} Searching the literature we found  that this modification  was first proposed by
 B.~S. Balakrishna, V. Sanyuk, J. Schechter, and A. Subbaraman, Phys. Rev. D 45, 344 (1992) (as a model of
the short-distance cutoff in QCD) but we are not aware of any further studies of this model.
\end{thebibliography}
\end{document}